\documentclass[10pt]{article}
\usepackage[utf8]{inputenc}
\usepackage{graphicx}
\usepackage[utf8]{inputenc}
\usepackage{amsmath}
\usepackage{amsfonts}
\usepackage{amssymb}
\usepackage{graphicx}
\usepackage{setspace}
\usepackage{authblk}
\usepackage{caption}
\usepackage{float}
\usepackage{cite}
\usepackage[left=2.0cm,right=2.0cm,top=2cm,bottom=2cm]{geometry}
\usepackage{color}
\usepackage{color,soul}
\usepackage{multicol}
\usepackage{subcaption}
\title{\textbf{Spectral Switch Anomalies in Sagnac Interferometer with respect to Galilean Frame}}

\author[1]{Shouvik Sadhukhan}
\author[2]{Saikat Sadhukhan}
\author[3]{Maruthi M Brundavanam}
\author[4]{C S Narayanamurthy}

\affil[1, 4]{\small{Department of Physics, Indian Institute of Space Science and Technology(IIST), P.O: Valiamala, Trivandrum - 695547, State: Kerala; India}}
\affil[3]{\small{Department of Physics, Indian Institute of Technology, Kharagpur-721302, West Bengal, India}}
\affil[4]{\small{Department of Mining Engineering, Indian Institute of
Engineering Science and Technology, Shibpur, Howrah-711 103, West
Bengal, India.}}
\affil[1]{\small{Email: shouvikphysics1996@gmail.com}}
\affil[3]{\small{Email: bmmanoj@phy.iitkgp.ac.in}}

\date{Dated : \today}
\begin{document}
\maketitle

\begin{abstract}
We report the spectral switch shift around spectral anomalies in a gyroscopic Sagnac interferometer which is normally used to calibrate the angular momentum of a gyroscope. The spectral shift in the rotating gyroscope is explained with respect to the longitudinal Doppler shift of the counter propagating beams in the Sagnac interferometer.\\

\end{abstract}

\begin{multicols}{2}
{

\section{Introduction}
Optical interferometric techniques are one of the most useful techniques in the measurements of classical dynamics and displacements. Optical gyroscope is one such instrument which is rigorously used in dynamical measurements \cite{32,33,34,23}. Its rotating property is used to measure the motion of air crafts as well as ships and submarines \cite{35,36,37,24,25,26}. Recently, optical gyroscope technique has been used in observational astronomy for astrophysical interferometric measurements \cite{34,27}, also the utility of gyroscope in gravitational wave measurement has become important due to its sensitivity and signal to noise ratio \cite{36,28}. Recently, some studies have been done with gyroscopic stage Sagnac interferometer where the rotational motion provided the necessary path difference for getting interference fringes \cite{29,30,31}. The counter propagating waves of such interferometer travels same path and hence external noises are reduced automatically. \par
In recent times, the spectral shifts around spectral switches are used to measure the Nano-displacement technique that is done with broad band polychromatic source in spectral interferometers \cite{1,2,3,38,39,40}. This Nano-displacement technique has not been used in the measurements of angular momentum of optical gyroscope. Also, the application of optical gyroscopic motion with white light can provide the results for Doppler shift. Hence, it is necessary to find the relation between the spectral shift around spectral switch and Doppler shift. Several experiments have been reported in literature using spectral shifts around spectral switches, for example, phase singularity in the complex analytic signal of a speckle pattern as indicators of displacement in nanometer scales \cite{4,5,6,7,8,41,42,43,44}, an interferometric phase-meter based on attenuated total internal reflection as nanoscale linear sensor \cite{9,10,11,12,13}, and spectral-domain phase microscopy for detection of nanometer-scale motions in living cells \cite{14,15,16,17,18}.\par
The usual gyroscopic sensitivity experiments are done using the interference fringe shift measurements. The monochromatic laser source is generally used here with fiber gyroscope. The fiber gyroscope are measured with a gyroscopic system including optical fibers and beam splitters in compact  structures. The optical fibers are generally taken as single mode fiber. The fringe shift measurements are not generally, too much efficient to provide the analysis of very small rotations. This is because the fringe shift becomes very small with small changes of angular speed as well as the radius. Hence, this problem can be solved with increase the sensitivity of the gyro sensor with polychromatic input. The polychromatic input can produce spectral switch and its shift around spectral anomalies under certain biasing conditions. This spectral switch shift can provide nano scale displacement measurements in linear moving operations. Hence, this mechanism is efficient to give very small scale changes in rotation and radius changes in gyroscope. The efficiency can also be modified or varied with changing the peak intensity and bandwidth of the input. This new technique have been represented in this present paper.\par
The spectral shifts around spectral switches are used to measure the angular momentum of the gyroscope based on modified Spectral Sagnac interferometer. The calibration curves for the measurements of angular momentum, radius and angular velocity of the gyroscope are obtained using analytical calculations and computer simulations. The results obtained in this paper will be useful in aircraft velocity measurements and Gravitational wave detection.\cite{19,20,21,22,24}\par
The paper has been organized as below. The Section 2 reports the mathematical overview of basic spectral interferometry with non-relativistic Sagnac interferometer along with the phase retrieval method. This formalism of Sagnac interferometer will be helpful to get the phases of our proposed experiment. Section 3 contains all results and analysis from our proposed experimental set -up. Finally, in section 4 we have provided detailed physical analysis of our results including Doppler shift and its calibrations.

\section{Spectral interferometric procedure and theory}
The proposed experimental geometry to elucidate our results is shown in Fig.1 for the spectral shift analysis. The Sagnac interferometer used in our analysis provides optical path difference due to Gyroscopic motion in a Galilean frame which in turn results in a spectral shift.  Fig.1 shows a typical Sagnac interferometer set-up where it,

\begin{figure}[H]
\centering
\begin{minipage}[b]{0.5\textwidth}
    \includegraphics[width=\textwidth]{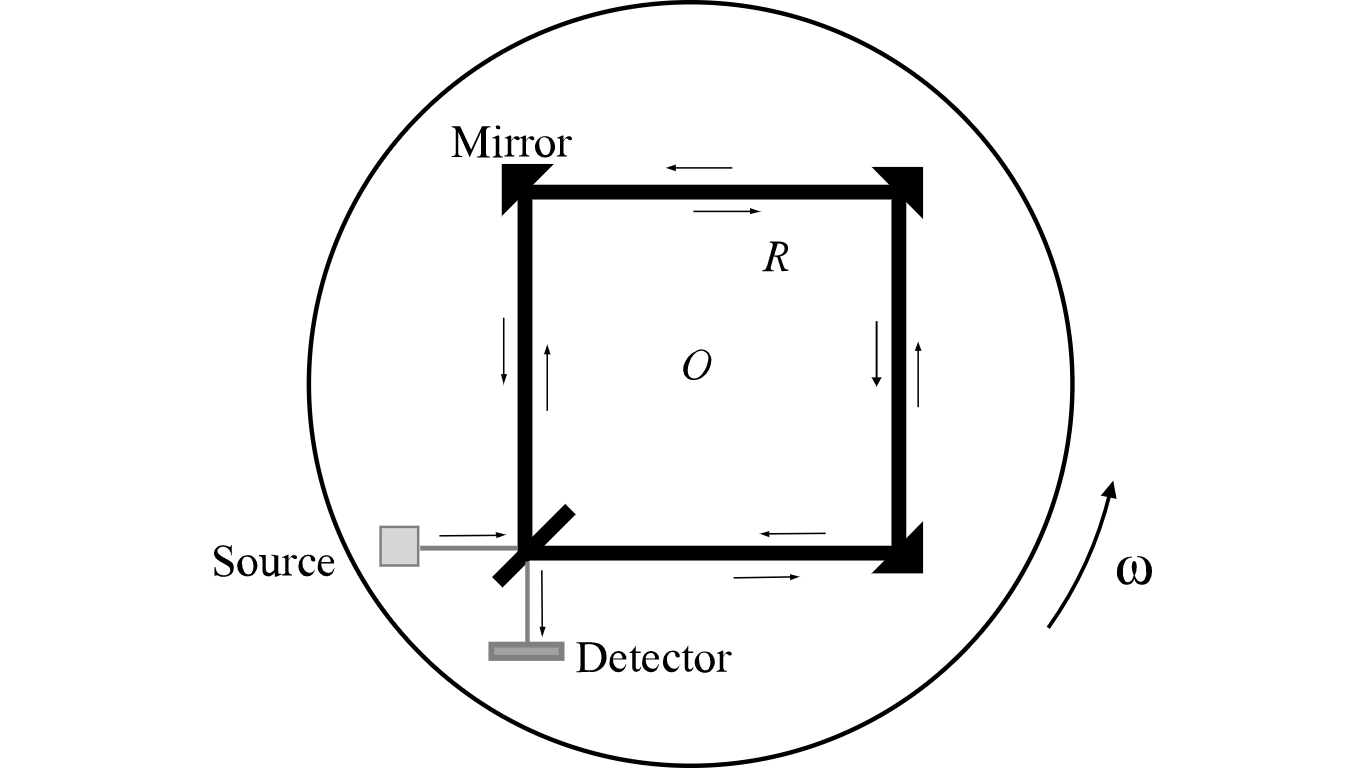}
    \caption{Sagnac Interferometer setup with the Gyroscopic table}
\end{minipage}
\end{figure}
consists of a dispersion-compensated Sagnac Interferometer illuminated with a white-light source and having a central wave length at 615 nm.  A fiber-coupled spectrometer is normally used at the output of the interferometer to measure the spectral modifications due to varying temporal correlation. We assume that the entire Sagnac interferometer set-up can be rotated non-relativistically \cite{2,3,4}. Here we also assume that the interferometer can have different angular motions in the table to calibrate the spectral switch with rotating motion. In such case since the angular speed is under Non-relativistic limit, we can assume that $\frac{\omega R}{c}\rightarrow 0$ and then the non-relativistic phase shift due to rotation in a Sagnac interferometer becomes,
\begin{equation}\label{10}
     \theta=\frac{8\pi^2}{c}\frac{R^2\omega}{\lambda}
\end{equation}
Therefore,the intensity can be written as follows.
\begin{equation}\label{11}
    I(\lambda)=H_0 \exp{(-\frac{1}{2}(\frac{\lambda-\lambda_0}{\Delta\lambda})^2)}(1+\cos{(\frac{8\pi^2}{c}\frac{R^2\omega}{\lambda})})
\end{equation}
This intensity function can provide the spectral switch and its shift with different values of angular rotation for non-relativistic motion.Consider $ I_{0}(\lambda) $ is the intensity of a Gaussian beam with central peak around the intensity at $H_0$ and the spectral domain intensity with a bandwidth of $\Delta\lambda $. 
\begin{equation}\label{1}
    I_0 (\lambda)=H_0 \exp{(-\frac{1}{2}(\frac{\lambda-\lambda_0}{\Delta\lambda})^2)}g(\lambda)
\end{equation}
where the term $g(\lambda)$ represents the assymmetry in the wave form. Then the function $ g(\lambda) $ is given by,
\begin{equation}\label{2}
\begin{split}
    g(\lambda)=1+\frac{A}{6}((\frac{\lambda-\lambda_0}{\Delta\lambda})^2-3(\frac{\lambda-\lambda_0}{\Delta\lambda}))\\-\frac{C}{24}((\frac{\lambda-\lambda_0}{\Delta\lambda})^4-6(\frac{\lambda-\lambda_0}{\Delta\lambda})^2+3)
    \end{split}
\end{equation}
In Eqn.2 the constant $A$ is the asymmetry factor and $C$ is the excess. The wave form should follow the following condition equal to,
\begin{center}
    $g(\lambda) \rightarrow
    \begin{cases}
      =1 & \text{for symmetric wave form spectrum} (A=B=0)\\
      \neq 1 & \text{for asymmetric wave form spectrum} (A\neq B\neq 0)
    \end{cases}$  
\end{center}
Since in our work, we have considered only the symmetric and non-excess singlet Gaussian Beam in spectral domain then $ A = B = 0$.\par 
If the input spectral domain field is $\varepsilon_0(\lambda)$, then the input intensity given in Eqn.1, can be rewritten as follows.
\begin{equation}\label{3}
    I_0(\lambda)=\varepsilon^{*}_0(\lambda)\varepsilon_0(\lambda)
\end{equation}
We have used the spectral domain of the spectrum in this present work. We have used the experimental geometry of Sagnac interferometer(Fig.1) and so the amplitudes on the two arms of the interferometer can be taken as $a(\nu)$ and $b(\nu)$ respectively. It can be noted here that both of them are dependent on the transmittance and reflectance of the Beam splitter. Assuming that the complex degree of coherence for this system is 1, the wave field for the two arms of the interferometer can be taken as equation 3 and 4($a(\nu)=rt\exp{(ikL_1)}$ and $b(\nu)=tr\exp{(ikL_2)}$). Now,the field amplitude along path 1 in the interferometer will be,
\begin{equation}\label{4}
   \varepsilon _1(\lambda)=a(\lambda)\varepsilon _{0}(\lambda)
\end{equation}
and for the path 2,
\begin{equation}\label{5}
   \varepsilon _2(\lambda)=b(\lambda)\varepsilon _{0}(\lambda)
\end{equation}
Then the total field after interference will be equal to,
\begin{equation}\label{6}
    \varepsilon (\lambda)=\varepsilon_1 (\lambda)+\varepsilon_2 (\lambda)
\end{equation}
Finally, the total intensity at the output can be found with $I(\lambda)=\varepsilon^{*} (\lambda) \varepsilon (\lambda)$. Then the spectral domain intensity output with arbitrary phase $\theta$, can be written as follows.\cite{2,3,4,24}
\begin{equation}\label{7}
     I(\lambda)=H_0 \exp{(-\frac{1}{2}(\frac{\lambda-\lambda_0}{\Delta\lambda})^2)}(1+\cos{(\theta)})
\end{equation}
Therefore, all these equations will be used to determine the interfered intensity spectrum. Since the square Sagnac interferometer has same arm lengths along both directions of the interfering beams,the phase difference comes only with the rotation of the entire Sagnac interferometer set up bench and so, we have considered $L_1=L_2$.

\subsection{Overview of Phase retrieval method}
In our present calculation we have used the predetermined phases induced in the interferometer arms due to the dispersive medium. But in our experimental set-up we have to introduce the phase retrieval method to get the phases with interpolation technique. We have used the following formulae for the phase retrieval technique.\cite{4,5,6}
\begin{itemize}
    \item [i] $I_{interference}(\lambda)=I(\lambda)(b(\lambda)+a(\lambda)+2\sqrt{b(\lambda)a(\lambda)}V(\lambda)\cos{(\theta)})$
    \item [ii] $VS(\lambda)=V(\lambda)\cos{(\theta)}$
    \item [iii] Interpolation $\rightarrow$ $\cos{(\theta)}=\frac{2VS(\lambda)-VS_{up}(\lambda)-VS_{down}(\lambda)}{VS_{up}(\lambda)-VS_{down}(\lambda)}$
\end{itemize}
 The above process is to use to retrieve the phases from interference to analyse the spectral shift with our results. For each speed of rotation we get the data for intensities and from those data we can use the above three points to derive the phases.

\section{Theoretical results under Galilean frame}
Initially we have fixed the radius and the angular speed of the gyroscopic table. After that we made some changes into the angular speed as well as the radius to find spectral shift in the spectral switch representation of our white light interferometry. Here, we first show the acceptable angular speed-radius doublets for getting first spectral switch in both frames. For getting spectral switch from Galilean frame experimental set-up, we have to follow the following equalities.
\begin{equation}
    \frac{8\pi^2}{c}\frac{R^2\omega}{\lambda_0}=\pi
\end{equation}
Hence, using this relation we can plot the necessary radius-angular speed doublets to get first spectral switch. Here, we considered a polychromatic input with peak wavelengths for non-relativistic as 615 nm. The Bandwidths of it has been taken as 238 nm .
\begin{figure}[H]
\centering
\begin{minipage}[b]{0.45\textwidth}
    \includegraphics[width=\textwidth]{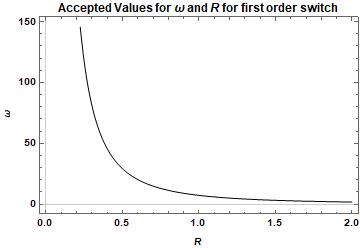}
    \caption{Accepted Values for $\omega$ and $R$ for first order switch in Non-Relativistic frame}
\end{minipage}
\end{figure}

Fig.2 shows the graph plotted between frequency $ \omega $ and $ R $, radius of Gyroscope (shown in Fig.1). This plot shows the accepted values required for $ \omega $ and radius $ R $ for first order switch in non-relativistic frame of reference. Then the plots shown in Fig 3a shows the graph plotted between wavelength $ \lambda $ and intensity $ I( \lambda) $ for $ R = 0.3048 m $ and with angular speeds 58.52 rps (Green), 58.82 rps (Black) and 59.12 rps (Red) respectively. These three angular velocities are providing different spectral switch. The radius-angular speed doublet $= (0.3048, 58.9)$ belongs to the trajectory with spectral switch position. Figures 3b, 3c and 3d show the enlarged part of total spectrum shown in Fig.3a.

}

\end{multicols}

\begin{figure}[H]
\centering
\begin{minipage}[b]{0.8\textwidth}
    \includegraphics[width=\textwidth]{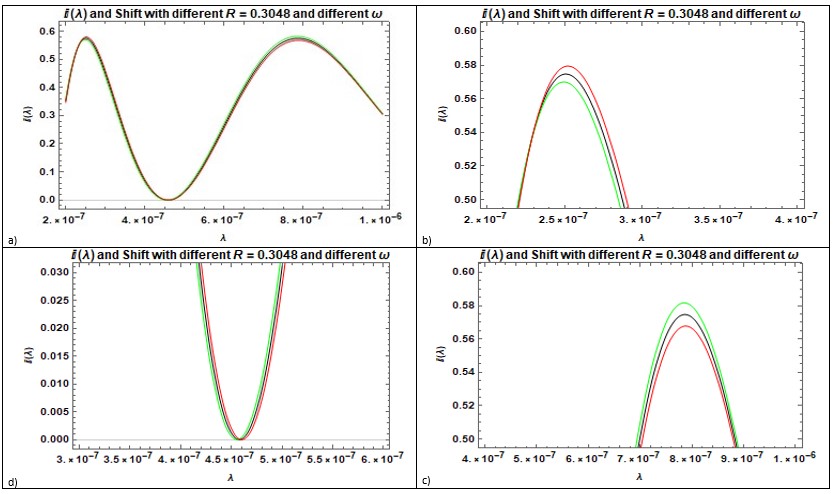}
    \caption{$I(\lambda)$ and Shift with $R = 0.3048 m$  for different $\omega$ i.e. $\omega= 58.52 rps$ (Green), $\omega= 58.82 rps$ (Black), $\omega= 59.12 rps$ (Red) (Non-Relativistic Case). Here Figure a) is the total spectrum and b), c) and d) are the zoomed parts of the spectrum.}
\end{minipage}
\end{figure}

\begin{multicols}{2}
{

In a similar way the plots shown in Fig.4a is between intensity $I(\lambda) $ and $ \lambda $ for angular speed of Gyroscopic stage $ \omega = 58.82 rps $ and for different $ R $ values. In Fig.4a, b, c and d show total spectrum and enlarged parts respectively. The spectral shifts are very evident from these plots. The radius $ R $ values chosen for these plots are equal to 0.3028 m (Green), 0.3048 m (Black) and 0.3068 (Red) respectively as these three radius valued systems are providing different spectral switch. Here also the radius-angular speed doublet $= (0.3048, 58.9)$ belongs to the trajectory with spectral switch position.

}

\end{multicols}

\begin{figure}[H]
\centering
\begin{minipage}[b]{0.8\textwidth}
    \includegraphics[width=\textwidth]{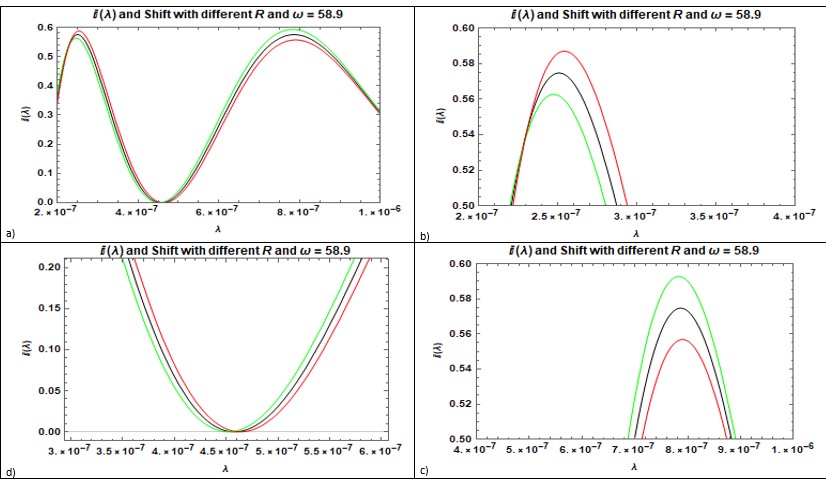}
    \caption{$I(\lambda)$ and Shift with $\omega = 58.82 rps$  for different $R$ i.e. $R= 0.3028 m$ (Green), $R= 0.3048 m$ (Black), $R= 0.3068 m$ (Red) (Non-Relativistic Case). Here Figure a) is the total spectrum and b), c) and d) are the zoomed parts of the spectrum.}
\end{minipage}
\end{figure}

\begin{multicols}{2}
{

Next, we analysed normalized spectral shift(NSS Shift). In this type of analysis for natural spectral shift, the variations considered for angular frequency $ \omega $ is from 0 rps to 300 rps for a fixed radii values equal to 0.83m and 0.3048m. The results are plotted in Fig.5 which is in non-relativistic domain and in case of Fig.6, we have used variation of $ R $ from 0.3 m to 2.5 m with fixed angular speed values of $\omega$ at 7.94 rps and 58.9 rps respectively.  The equation for getting NSS used in this analysis is given by $ NSS=\frac{\lambda_0-\lambda}{\lambda_0} $ Here, we have also provided the peak values of the spectral Intensity with path delay due to different $ \omega $ values and $ R $ values respectively.Further Figures 7 and 8 show the plots of peak spectral intensities with respect to various angular speeds. In the graph shown in Fig.7 two graphs are plotted for peak spectral intensity versus angular speed ranges from i.e $ \omega $ varying from $ 0 $ to $ 300  rps $ for $ R = 0.83 m $ $ 0.3048 m $ respectively. On the other hand, in  case of graphs plotted in Fig.8 the peak spectral intensity is plotted with respect to $ R $ values $ 0.3 m $ and $ 0.2.5 m $ for two different angular speed values $ \omega = 7.94 rps $ and $ \omega = 58.9 rps $ respectively. In both cases one can see the clear spectral shifts.  \par
Another important observation is related to spectrum switch amplitudes. This can be obtained from maximum and minimum values of natural spectral shift values. The variation of spectral switch amplitude values from natural spectral shift values are plotted in Fig. 9 and Fig.10 respectively. The spectral switch amplitude value can be calculated from $SSA=\frac{\lambda_{red}-\lambda_{blue}}{\lambda_0}$. The $ \lambda_{red} $ and $ \lambda_{blue} $ are the wavelengths of red and blue shifted peaks values across the spectral switch. Fig.9 shows the graph plotted between SSA values in y-axis and $ \Delta \omega $ values in x-axis for $ R = 0.83m $ and $ R = 0.3048m $ values respectively. Similarly Fig.10 shows the graph plotted between SSA values in y-axis and $ \Delta R $ values in x-axis for two different angular speed values $ \omega = 7.94 rps$ and $ 58.9 rps $ respectively. From these two plots one can clearly see the changes in spectral shift amplitude values when angular speed and radius of Gyroscope is changed.\par
Then, in Fig.11 we have shown the changes obtained in normalized spectral shift (NSS) when optical path lengths are changed for 11 peak wavelengths. The wavelengths values range from $ 525 nm $ to $ 725 nm $. It is very clear from the graph shown in Fig.11 that NSS values change for each wavelengths and they are not same even for two neighbouring wavelengths. 

}

\end{multicols}

\begin{figure}[H]
\begin{subfigure}{.33\textwidth}
  \centering
  \includegraphics[width=\textwidth]{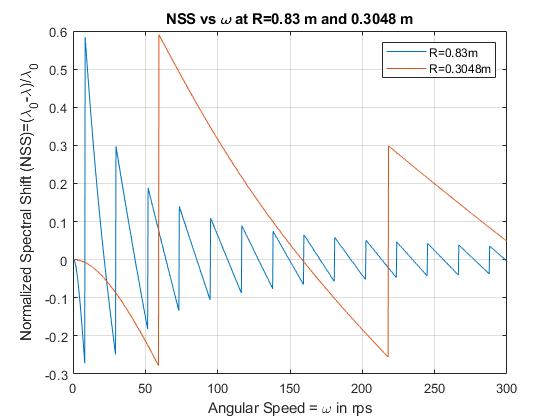}
    \caption{\small{NSS vs $\omega$}}
  \label{fig:sub-first}
\end{subfigure}
\begin{subfigure}{.33\textwidth}
  \centering
  \includegraphics[width=\textwidth]{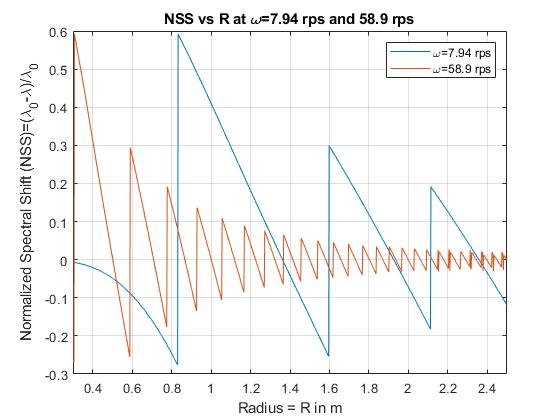}
    \caption{\small{NSS vs $R$}}
  \label{fig:sub-second}
\end{subfigure}
\begin{subfigure}{.33\textwidth}
  \centering
  \includegraphics[width=\textwidth]{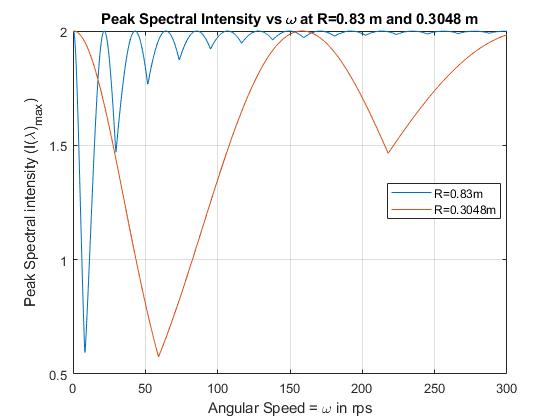}
    \caption{\small{Peak Spectral Intensity vs $\omega$}}
  \label{fig:sub-third}
\end{subfigure}


\begin{subfigure}{.33\textwidth}
  \centering
  \includegraphics[width=\textwidth]{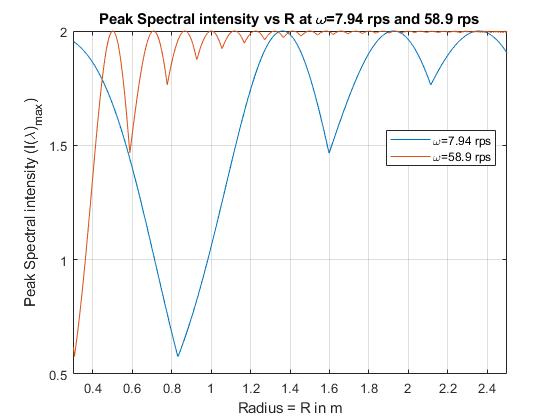}
    \caption{\small{Peak Spectral Intensity vs $R$}}
  \label{fig:sub-fourth}
\end{subfigure}
\begin{subfigure}{.33\textwidth}
  \centering
  \includegraphics[width=\textwidth]{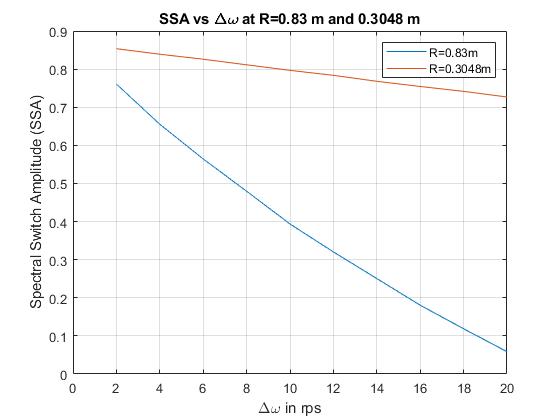}
    \caption{\small{SSA vs $\Delta\omega$}}
  \label{fig:sub-fifth}
\end{subfigure}
\begin{subfigure}{.33\textwidth}
  \centering
  \includegraphics[width=\textwidth]{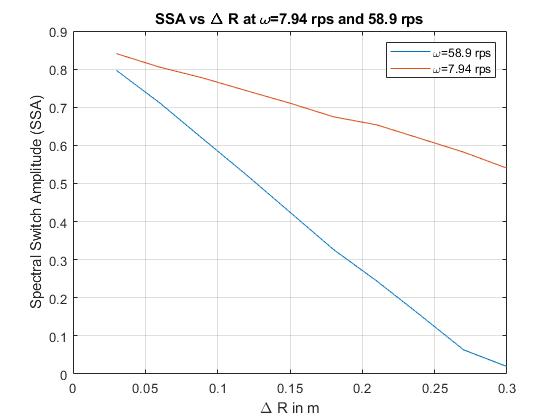}
    \caption{\small{NSS vs $\Delta R$}}
  \label{fig:sub-sixth}
\end{subfigure}

\caption{Black for Radiation and WDM, Blue for CDM, Red for Quintessence, Pink for Quintom and Purple for Phantom universe. Fig a), c) and e) are with $\omega\in[0,300]rps$ and fixed radii $R=0.83 m$ and $R=0.3048 m$ (Non-Relativistic Case) and the fig. b), d) and f) are with $\Delta R\in[0.03,0.3]m$ and fixed angular speed $\omega=7.94 rps$ and $\omega=58.9 rps$ (Non-Relativistic Case)}
\label{fig:fig}
\end{figure}

\begin{multicols}{2}
{

\begin{figure}[H]
\centering
\begin{minipage}[b]{0.45\textwidth}
    \includegraphics[width=\textwidth]{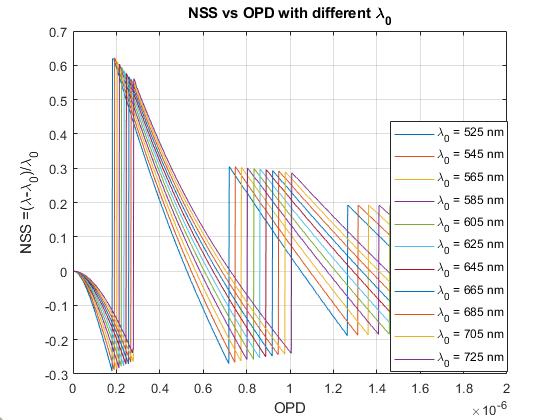}
    \caption{Normalized Spectral Shift (NSS) vs Optical Path Difference (OPD) with different peak wavelengths as $\lambda_0\in [525,725] nm$ with $\Delta\lambda_0 = 20 nm$}
\end{minipage}
\end{figure} 

\section{Physical interpretation of results}
In figure 11 we have discussed the NSS shift with different input peak wavelengths values of the broadband white light. It is observed that the number of switches are decreasing for a certain fixed OPD range with increase in peak wavelengths. Similar kind of NSS shift and decrease in the number of switches are happening in the graphs shown in Fig 5 and Fig.6 respectively. In this analysis of Fig. 5 and Fig.6 we have kept the peak wavelength fixed at $\lambda_0 = 615 nm$. Thus it can be concluded that there must have some wavelength change before the interference due to rotation in Gyroscope.\par
This phenomena is similar to the Doppler shift of colour shift. This is because, the velocity and radius changes provide Doppler shifts in each wavelengths along with the peak value. The waves moving in same direction of motion are getting red shifted and the waves opposite to it are getting blue shifted. Hence, before interference, there must have some wavelength shift of Doppler shift which again modifies the intensity values. That's how one can measure the Doppler shift with the measurement of spectral switch shift. On the other hand, in our work the Doppler shift is happening not only due to the presence of changing velocity or frame movement, but also it is happening due to change in angular momentum which is coming due to the change in radius. The detail calculation of Doppler shift and the effect of angular momentum is derived below. The Doppler shift for non-relativistic speed can be represented as follows. ($c$ is the speed of light)
\begin{equation}
    z=\frac{\triangledown\lambda}{\lambda}=\frac{\omega R}{c}
\end{equation}
where we have the shifted wavelength as follows.
\begin{equation}
    \lambda_{changed}=\lambda_{initial}(1+z)=\lambda_{initial}(1+\frac{\omega R}{c})
\end{equation}
As we are using polychromatic light source, all the wavelengths should be shifted before interference in Sagnac interferometer. On the other hand, in our case, the phase difference due to rotation can be substituted with the Doppler wavelength shift. Then the final functional form of the spectral intensity will become,
\begin{equation}
\begin{split}
    I(\lambda)=H_0 \exp{(-\frac{1}{2}(\frac{\lambda-\lambda_0}{\Delta\lambda})^2(1+\frac{R\omega}{c})^2)}\\\times(1+\cos{(\frac{8\pi^2 R}{\lambda}\frac{\triangledown\lambda}{\lambda})})
    \end{split}
\end{equation}
This equation can also be written as follows.
\begin{equation}
\begin{split}
    I(\lambda)=H_0 \exp{(-\frac{1}{2}(\frac{\lambda-\lambda_0}{\Delta\lambda})^2(1+\frac{R\omega}{c})^2)}\\\times(1+\cos{(\frac{8\pi^2 c }{\omega\lambda}(\frac{\triangledown\lambda}{\lambda})^2)})
    \end{split}
\end{equation}
Here the term $(1+\frac{R\omega}{c})^2$ is a modification factor for Doppler shifted interference intensity in non-relativistic frame. For non-relativistic case we must have $R\omega<<c$. Hence the modification factor used here, can be assumed to be $\approx 1$ and that's why we can find negligible difference of intensity with earlier results. Hence, using those relations we can discuss the spectral switch shift with respect to either Doppler shift i.e. $z=\frac{\triangledown\lambda}{\lambda}$ or colour shift ($\lambda-$shift) i.e. $\triangledown\lambda$. Now we can represent the NSS shift and peak spectral intensity value with changing $z$ as well as $\triangledown\lambda$. We can also represent the calibration for the measurements of Doppler shift and colour shift using spectral switch amplitude. The Eqn.15 provides the proof for Doppler shift due to change of angular velocity where as the Eqn.16 provides the proof of Doppler shift due to change of radius. Hence, to calibrate the measurements of Doppler shift we must have to fix either radius or angular velocity at constant values.

\begin{figure}[H]
\centering
\begin{minipage}[b]{0.45\textwidth}
    \includegraphics[width=\textwidth]{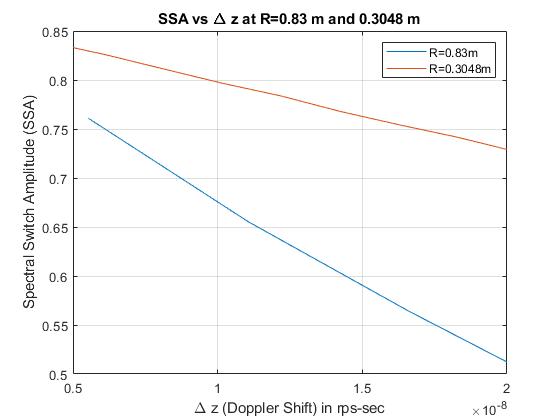}
    \caption{SSA vs $\Delta z$ with $\Delta z\in[5,20] nm $ and fixed radii $R=0.83 m$ and $R=0.3048 m$ (Non-Relativistic Case)}
\end{minipage}
\end{figure} 

\begin{figure}[H]
\centering
\begin{minipage}[b]{0.45\textwidth}
    \includegraphics[width=\textwidth]{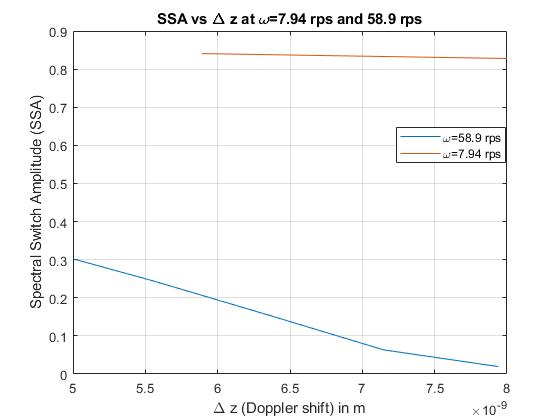}
    \caption{SSA vs $\Delta z$ with $\Delta z\in[5,8] nm $ and fixed angular speed $\omega=7.94 rps$ and $\omega=58.9 rps$ (Non-Relativistic Case)}
\end{minipage}
\end{figure}

\section{Conclusion}

In this work we have tried to introduce the idea of spectral switch and its shifting in Sagnac interferometric setup. The principle speciality of this specific work is that we have established the spectral switch interferometric optical sensing technique in presence of gyroscopic motion. The gyroscopic motion is very much useful in observational astronomy and our work may interpret the spectral switch optical sensing in it. Here we have also incorporated the physical analysis with doppler shift and spectral switch shift.\par
We have discussed our results depending upon an experimental set up with white light source and gyroscopic stage interferometer with non relativistic speed. The results have been simulated and discussed the spectral amplitude shift accordingly. In each cases we have mentioned two types of results, one for fixed radius and another for fixed angular speed. The Doppler shift has been also discussed for those two cases. We have observed that for the measurements of Doppler shift, fixed radius technique is less sensitive than fixed angular speed one.\par
From the NSS figures we can conclude that the sensitivity of our experiment should decrease with decrease of radius as because for higher radii, we can get large change for small phase difference. On the other hand, the sensitivity also increases with increase in angular speed of the gyroscopic stage.\par
The results found in this work can be established in realistic experiments. The gyrosencors are well established sensitivity technologies available in literature. The gyro fiber sensors are included in those extensively used gyrosensing operations available in literatures. Hence, we can introduce the fiber optical gyroscope into polychromatic input measurements as discussed in this manuscript. This gyro fiber technology can reduce biasing the radius and angular momentum to find the spectral switch point to make the sensor.

\section*{Acknowledgement}
Shouvik Sadhukhan and C S Narayanamurthy Acknowledge the SERB/DST(Govt. Of India) for providing financial support via the  project grant CRG/2020/003338 to carry out this work.

\section*{Disclosures}
The authors declare no conflicts of interest.

\section*{Data availability}
No data were generated or analyzed in the presented research.

}

\end{multicols}

\end{document}